\documentclass[letterpaper,english,10pt,aps,letter,superscriptaddress,prl,twocolumn]{revtex4}
\pdfoutput=1
\usepackage[T1]{fontenc}
\usepackage[latin1]{inputenc}
\usepackage{xcolor}
\usepackage{pdfcolmk}
\usepackage{verbatim}
\usepackage{graphicx}
\usepackage{amsmath, amsthm, amssymb}
\usepackage{amsfonts}

\pdfpageheight\paperheight
\pdfpagewidth\paperwidth

\renewcommand{\citet}[1]{\cite{#1}}
\renewcommand{\eqref}[1]{(\ref{#1})}
\newcommand{\revision}[1]{#1}

\begin{document}

\title{Reversible Diffusion by Thermal Fluctuations}

\author{Aleksandar Donev, Thomas G. Fai and Eric Vanden-Eijnden}
\affiliation{Courant Institute of Mathematical Sciences, New York University,
New York, NY 10012}

\begin{abstract}
A model for diffusion in liquids that couples the dynamics of tracer
particles to a fluctuating Stokes equation for the fluid is
investigated in the limit of large Schmidt number.  In this limit,
the concentration of tracers is shown to satisfy a closed-form
stochastic advection-diffusion equation that is used to investigate
the collective diffusion of hydrodynamically-correlated tracers
through a combination of Eulerian and Lagrangian numerical methods.
This analysis indicates that transport in liquids is quite distinct
from the traditional Fickian picture of diffusion. While the
ensemble-averaged concentration follows Fick's law with a diffusion
coefficient that obeys the Stokes-Einstein relation, each instance
of the diffusive mixing process exhibits long-ranged giant
fluctuations around its average behavior. We construct a
class of mesoscopic models for diffusion in liquids at different
observation scales in which the renormalized diffusion coefficient
depends on this scale. This indicates that the Fickian
diffusion coefficient in liquids is not a material constant, but
rather, changes with the scale at which experimental measurements are performed. 
\end{abstract}

\maketitle

\global\long\def\V#1{\boldsymbol{#1}}
 \global\long\def\M#1{\boldsymbol{#1}}
 \global\long\def\Set#1{\mathbb{#1}}

\global\long\def\D#1{\Delta#1}
 \global\long\def\d#1{\delta#1}

\global\long\def\norm#1{\left\Vert #1\right\Vert }
 \global\long\def\abs#1{\left|#1\right|}

\global\long\def\grad{\boldsymbol{\nabla}}
 \global\long\def\av#1{\langle#1\rangle}

\setlength{\abovedisplayskip}{0.4ex}\setlength{\belowdisplayskip}{0.4ex}
\setlength{\abovedisplayshortskip}{0.15ex}\setlength{\belowdisplayshortskip}{0.15ex}

Diffusion is one of the most ubiquitous transport process.  It is,
arguably, the simplest dissipative mechanism. Fick's law of diffusion
is ``derived'' in most elementary textbooks, and relates diffusive
fluxes to the gradient of chemical potentials via a diffusion
coefficient that is typically thought of as a material property. Yet,
there are several hints that diffusion in liquids is, in fact, rather
subtle. A first hint is that the Stokes-Einstein (SE) prediction for
the diffusion coefficient involves the viscosity of the fluid, a
seemingly independent transport property. This suggests a connection
between momentum transport and diffusion and may explain why the SE
prediction is in surprisingly reasonable agreement with measurements
even in cases where it should not apply at all, such as molecular
diffusion. A second hint is that nonequilibrium diffusive transport is
accompanied by ``giant'' long-range correlated thermal fluctuations
\citet{LongRangeCorrelations_MD,GiantFluctuations_Universal,%
  FluctHydroNonEq_Book}, which have been measured using light
scattering and shadowgraphy techniques
\citet{GiantFluctuations_Nature,GiantFluctuations_Universal,%
  GiantFluctuations_Cannell,FractalDiffusion_Microgravity}.  It is now
well-understood that these unexpected features of diffusion in liquids
stem from the contributions of advection by thermal velocity
fluctuations \citet{DiffusionRenormalization_I,%
  ExtraDiffusion_Vailati,DiffusionRenormalization,Nanopore_Fluctuations}.
It has long been appreciated in statistical mechanics and
nonequilibrium thermodynamics circles that thermal fluctuations
exhibit long-ranged correlations in nonequilibrium settings
\citet{LongRangeCorrelations_MD,FluctHydroNonEq_Book}.  The aim of
this Letter is to show that these fluctuations are also of overarching
importance to transport in fluids, a fact that has not been widely
recognized so far.

In either gases, liquids or solids, one can, at least in principle,
coarse-grain Hamiltonian dynamics for the atoms (at the classical
level) to obtain a model of diffusive mass transport at hydrodynamic
scales. This procedure is greatly simplified by first coarse-graining
the microscopic dynamics to a simpler stochastic description, which is
done by using kinetic theory for gases or Markov jump models for
diffusion in solids. In both cases the picture that emerges is that of
independent Brownian walkers performing uncorrelated random walks in
continuum (gases) or on a lattice (solids). By contrast, in liquids
the physical picture is rather different and must account for
hydrodynamic correlations among the diffusing particles. In a liquid,
molecules become trapped (caged) over long periods of time, as they
collide with their neighbors. Therefore, momentum and energy are
exchanged (diffuse) much faster than the molecules themselves can
escape their cage. The main mechanism by which molecules diffuse is
the motion of the whole cage when a large-scale velocity fluctuation
(coordinated motion of parcels of fluid) moves a group of molecules
and shifts and rearranges the cage.

Most previous theoretical studies of molecular diffusion are based on
some form of mode-mode coupling, which is essentially a perturbative
analysis in the strength of the thermal fluctuations
\citet{DiffusionRenormalization_I,DiffusionRenormalization_II,%
  DiffusionRenormalization_III,ExtraDiffusion_Vailati,%
  Nanopore_Fluctuations,DiffusionRenormalization}.  In this Letter we
formulate a simple model for diffusion in liquids at microscopic and
mesoscopic scales and use it to make a precise assessment of the
contribution of fluctuations to diffusive transport.  Our model is a
simplified (coarse-grained) representation of the complex molecular
processes that underlie mass transport in liquids. It mimics all of
the crucial features of realistic liquids, while also being tractable
analytically and numerically.  Through a mix of theoretical and
numerical studies, we show that this model exhibits realistic physical
behaviors that differ from those of standard models of Fickian
diffusion via uncorrelated random walks.  In particular, we find that
there is an unexpected connection between flows at small and large
scales, and at microscopic and mesoscopic scales diffusion in liquids
resembles turbulent diffusion.

Our model describes the motion of passive tracer particles advected
by thermal velocity fluctuations, and can be used to describe the dynamics of
fluorescently-labeled molecules in a Fluorescence Recovery After
Photo-bleaching (FRAP) experiment, the transport of nano-colloidal
particles in a nanofluid, or the motion of the molecules in a simple
fluid. \revision{We will neglect direct interactions among the particles, which
is appropriate when tracers are dilute.}
The evolution of the incompressible fluid velocity, $\V
v\left(\V r,t\right)$ with $\grad\cdot\V v=0$, is assumed to satisfy
the linearized fluctuating Navier-Stokes equation
\begin{equation}
\rho\partial_{t}\V v+\grad\pi=\eta\grad^{2}\V v+\sqrt{2\eta k_{B}T}\,\grad\cdot\M{\mathcal{W}},\label{eq:v_eq}
\end{equation}
where $\M{\mathcal{W}}\left(\V r,t\right)$ denotes a white-noise
symmetric tensor field (stochastic momentum flux) with \revision{covariance
chosen to obey a fluctuation-dissipation principle
\citet{FluctHydroNonEq_Book},
\begin{equation}
\av{\mathcal{W}_{ij}(\V r,t)\mathcal{W}_{kl}(\V r^{\prime},t^{\prime})}=\left(\delta_{ik}\delta_{jl}+\delta_{il}\delta_{jk}\right)\delta(t-t^{\prime})\delta(\V r-\V r^{\prime}).
\end{equation}
}

The details of the microscopic coupling
between the fluid and the passive tracer are complicated
\citet{StokesEinstein_BCs} and some approximations are required.  We
will assume that the evolution of the concentration of a large
collection of tracers, $c\left(\V r,t\right)$, can be modeled via a
fluctuating advection-diffusion equation
\begin{equation}
\partial_{t}c=-\V u\cdot\grad c+\chi_{0}\grad^{2}c.\label{eq:c_eq_original}
\end{equation} 
where $\chi_{0}$ is the \emph{bare} (molecular) diffusion coefficient
and the advecting velocity $\V u$ is obtained by convolving the fluid
velocity $\V v$ with a smoothing kernel ${\M{K}_{\sigma}}$ that
filters out features below the molecular scale $\sigma$,
\begin{equation}
\V u\left(\V r,t\right)=\int{\M{K}_{\sigma}}\left(\V r,\V r^{\prime}\right)\V v\left(\V r^{\prime},t\right)d\V r^{\prime}\equiv \left( {\M{K}_{\sigma}}\star\V v \right) \left(\V r,t\right),
\end{equation}
and preserves the zero-divergence condition, $\grad\cdot\V u=0$.  
\revision{Physically, one can think of $\sigma$ as representing
   the size of the molecular cage
   in the case of molecular diffusion and the
radius of the tracer particles for colloidal diffusion.} We
stress that the smoothing of the fluctuating velocity field $\V v$ is
necessary to avoid divergence (ultraviolet catastrophe) of the
effective diffusion coefficient of the tracer particle obtained
below. Thus, the molecular scale details
enter~\eqref{eq:c_eq_original} in two ways: through the term
$\chi_{0}\grad^{2}c$ \textit{and} the smoothing of $\V u$ below scale
$\sigma$. As we will see below, the smoothing turns out to be more
important for transport than the first. Molecular dynamics
simulations have confirmed that \eqref{eq:v_eq} and
\eqref{eq:c_eq_original} accurately model diffusive mixing between two
initially phase-separated fluids down to essentially molecular scales
\citet{LowMachExplicit}. % Here we will assume that the
% concentration field varies at length scales much larger than the
% typical distance between tracers, allowing for a meaningful
% definition of a smooth concentration field $c\left(\V r,t\right)$; a
% more detailed model are presented elsewhere
% \citet{DiffusionJSTAT}.
\revision{Note that an additional (mathematically problematic)
   multiplicative noise term $\grad\cdot\left(\sqrt{2\chi_{0}c}\,\M{\mathcal{W}}_{c}\right)$,
   where $\M{\mathcal{W}}_{c}(\V r,t)$ is a white noise random vector field,
needs to be included in (\ref{eq:c_eq_original}) to capture equilibrium concentration
fluctuations \cite{DiffusionJSTAT}; we do not include this term in this work in order to focus our attention on the
nonequilibrium (giant) fluctuations that appear due to the advection by the fluctuating velocity.}

%\section{The Limit of Large Schmidt Number}

In liquids, diffusion of mass is much slower than that of momentum,
i.e. the velocity evolves fast compared to the concentration. This
separation of time scales is measured by the Schmidt number and it can
be used to eliminate the velocity
\citet{ModeElimination_Papanicolaou,ModeElimination_MTV}. This
procedure, the details of which are presented elsewhere
\citet{DiffusionJSTAT}, gives a \emph{limiting} stochastic
advection-diffusion equation for the concentration which reads
\footnote{%
  Here $\V w\odot\grad c$ and $\V w\cdot\grad c$ are short-hand
  notations for $\sum_{k}\left(\V{\phi}_{k}\cdot\grad c\right)\circ
  d\mathcal{B}_{k}/dt$ and $\sum_{k}\left(\V{\phi}_{k}\cdot\grad
    c\right) d\mathcal{B}_{k}/dt$, respectively, where
  $\mathcal{B}_{k}\left(t\right)$ are independent Wiener processes and
  $\V{\phi}_{k}$ are basis functions such that
  $\M{\mathcal{R}}\left(\V r,\V
    r^{\prime}\right)=\sum_{k}\V{\phi}_{k}\left(\V r\right) \otimes
  \V{\phi}_{k}\left(\V r^{\prime}\right)$.}
\begin{equation}
  \label{eq:limiting_Strato_Ito}
  \begin{aligned}
    \partial_{t}c&=-\V w\odot\grad
    c+\chi_{0}\grad^{2}c &\text{(S)}\\
    & =-\V w\cdot\grad c+\chi_{0}\grad^{2}c
    +\grad\cdot\left[\M{\chi}\left(\V r\right)\grad c\right]  &\text{(I)}
  \end{aligned}
\end{equation}
where the first equality shows the equation in Stratonovich's
interpretation and the second in Ito's.  Here the advection velocity
$\V w\left(\V r,t\right)$ is divergence free ($\nabla \cdot \V w =0$) and
white-in-time, with covariance proportional to a Green-Kubo integral
of the auto-correlation function of $\V u(\V r,t)$, i.e. $\av{\V
  w\left(\V r,t\right)\otimes\V w\left(\V
    r^{\prime},t^{\prime}\right)}= \M{\mathcal{R}}\left(\V r,\V r^{\prime}\right)
  \delta\left(t-t^{\prime}\right)$ where
\begin{equation}
  \M{\mathcal{R}}\left(\V r,\V r^{\prime}\right)=
  2\int_{0}^{\infty}\av{\V u\left(\V r,t\right)\otimes
    \V u\left(\V r^{\prime},t+t^{\prime}\right)}dt^{\prime},\label{eq:R_r}
\end{equation}
and the enhancement of the diffusion coefficient is 
$\M{\chi}\left(\V r\right)=\tfrac{1}{2}\M{\mathcal{R}}\left(\V r,\V r\right)$.
%=\int_{0}^{\infty}\av{\V u\left(\V r,t\right)\otimes\V u\left(\V
%r,t+t^{\prime}\right)}dt^{\prime}.
%\label{eq:chi_r_general}
Similar equations, but with a distinct form of the covariance
$\M{\mathcal{R}}$, appear in the Kraichnan model of turbulent
transport \citet{Kraichnan_SelfSimilar,Kraichnan_SelfSimilarPRL}
(see Sec.~4.1 in \citet{TurbulenceClosures_Majda}).
\revision{It can be shown \citet{DiffusionJSTAT} that at the Lagrangian level
(individual tracer trajectories) \eqref{eq:limiting_Strato_Ito} is equivalent
to the well-known equations of Brownian Dynamics  with
hydrodynamic interactions (correlations)
of a form similar to the Rotne-Prager tensor \citet{BrownianBlobs}, which
is widely used as a model for diffusion in dilute colloidal suspensions.}

Equation~\eqref{eq:limiting_Strato_Ito} has properties that may seem
paradoxical at first sight but have important implications for
transport in liquids. Indeed notice that it is easy to take the
average of this equation in Ito's form to deduce that the ensemble
average of the concentration obeys Fick's law,
\begin{equation}
  \partial_{t}\av c=\grad\cdot\left(\M{\chi}_{\text{eff}}\grad\av
    c\right) \quad\text{where}\ \ \M{\chi}_{\text{eff}}
  =\chi_{0}\M{I}+\M{\chi},
  \label{eq:dc_dt_mean}
\end{equation}
which is a well-known result that can be justified rigorously (c.f.
Eq.~(255) in \citet{ModeElimination_MTV}) and holds even in the
absence of bare diffusion, $\chi_{0}=0$. This is surprising
considering that \eqref{eq:limiting_Strato_Ito} is time-reversible
when $\chi_0=0$, as made clear by Stratonovich's form of this equation.
Furthermore, the same equation \eqref{eq:dc_dt_mean} holds for
\emph{all} moments of~$c$ when $\chi_0=0$. This is no contradiction:
the ``dissipative'' term $\grad\cdot\left[\M{\chi}\left(\V
    r\right)\grad c\right]$ and the stochastic forcing term $-\V
w\cdot\grad c$ are signatures of the same physical process, advection
by thermal velocity fluctuations. Including the first term but
omitting the second violates fluctuation-dissipation balance and
\emph{cannot} be justified. For example, the stochastic terms in
\eqref{eq:limiting_Strato_Ito} need to be retained to obtain the giant
fluctuations seen in a \emph{particular instance} (realization) of the
diffusive mixing process.

%\subsection{Connections to the Stokes-Einstein Relation}

Next we estimate the diffusion enhancement, $\M{\chi}(\V r) = \tfrac12
\M{\mathcal{R}}(\V r,\V r) $, to get an intuitive understanding of its
role. Since $\V v\left(\V r,t\right)$ solves the linearized
fluctuating Navier-Stokes equation \eqref{eq:v_eq}, it is not hard to
show that
\begin{equation}
\int_{0}^{\infty}\av{\V v\left(\V r,t\right)\otimes\V v\left(\V r^{\prime},t+t^{\prime}\right)}dt^{\prime}=\frac{k_{B}T}{\eta}\M G\left(\V r,\V r^{\prime}\right),\label{eq:v_cov_Stokes}
\end{equation}
where $\M G$ is the Green's function (Oseen tensor) for the steady
Stokes equation with unit viscosity, $\grad\pi=\grad^{2}\V v+\V f$
subject to $\grad\cdot\V v=0$ and appropriate boundary conditions.
Inserting this expression in~\eqref{eq:R_r} implies that
\begin{equation}
  \M{\mathcal{R}}\left(\V r,\V r^{\prime}\right) 
  = \left(\M{K}_{\sigma}\star\M G\star\M{K}^T_{\sigma}\right)\left(\V r,\V r^{\prime}\right)
%\M{\chi})(=\frac{k_{B}T}{\eta}\int{\M{K}_{\sigma}}\left(\V r,\V r^{\prime}\right) \M G\left(\V r^{\prime},\V r^{\prime\prime}\right){\M{K}^T_{\sigma}}\left(\V r,\V r^{\prime\prime}\right)d\V r^{\prime}d\V r^{\prime\prime}.
\label{eq:chi_r_Stokes}
\end{equation}
% A sample of the Brownian increment $\sum_{k}\V{\phi}_{k}d\mathcal{B}_{k}$
% can be obtained by solving a steady Stokes problem with a suitable
% random forcing (fluctuating stress), and then convolving the velocity
% with the filter ${\M{K}_{\sigma}}$.
To proceed, recall that for an infinite isotropic system, $\M{G}(\V
r,\V r^{\prime}) \equiv \M{G}(\V r-\V r^{\prime})$ is the Oseen tensor, the Fourier
tranform of which reads $\widehat{\M G}_{\V k}=k^{-2}\left(\M
  I-k^{-2}\V k\otimes\V k\right)$. Let us employ an isotropic
filtering kernel ${\M{K}_{\sigma}}$ that cuts off the fluctuations in
the advective velocity $\V w$ at \emph{both} large and small scales to
account respectively for the finite extent of the system $L$ and the
filtering at the molecular scale $\sigma$, and assume that the Fourier transform of 
$\M{\mathcal{R}}(\V r-\V r^{\prime}) \equiv \M{\mathcal{R}}(\V r,\V r^{\prime})$ is
\begin{equation}
  \Hat{\M{\mathcal{R}}}_{\V k}=\frac{2 k_{B}T}{\eta}
  \frac{k^{2}L^{4}}{\left(1+k^{4}L^{4}\right)
    \left(1+k^{2}\sigma^{2}\right)}
  \left(\M I-\frac{\V k\otimes\V k}{k^{2}}\right)
  \label{eq:chi_k_filtered}
\end{equation}

Converting \eqref{eq:chi_k_filtered} to real space gives an isotropic
enhancement to the diffusion tensor
$\M{\chi}=\M{\mathcal{R}}(0)/2=\left(2\pi\right)^{-d}\int\left(\hat{\M{\mathcal{R}}}_{\V k}/2\right)d\V k=\chi\M I$.
This Fourier integral is exactly the one that appears in
the linearized steady-state (static) approximate renormalization
theory when $\nu\gg\chi_{0}$ \citet{DiffusionRenormalization_I,
  ExtraDiffusion_Vailati, DiffusionRenormalization}.  Here we obtain
the same result with a simple, general, and precise calculation that
gives~\footnote{%
  Some of the coefficients in \eqref{eq:chi_SE} depend on the exact
  form of the spectrum $\hat{\M{\mathcal{R}}}_{\V k}$
  in~\eqref{eq:chi_k_filtered}.}
for $L\gg\sigma$
\begin{equation}
\chi\sim\frac{k_{B}T}{\eta}\begin{cases}
\left(4\pi\right)^{-1}\ln(L/\sigma) & \mbox{if }d=2\\
\left(6\pi\sigma\right)^{-1} & \mbox{if }d=3.
\end{cases}\label{eq:chi_SE}
\end{equation}
In three dimensions \eqref{eq:chi_SE}
gives the Stokes-Einstein prediction
$\chi\sim\chi_{SE}=k_{B}T/\left(6\pi\eta\sigma\right)$ for the
diffusion coefficient of a slowly-diffusing no-slip rigid sphere of
radius $\sigma$. In two dimensions, the effective diffusion
coefficient grows logarithmically with system size, in agreement with
the Einstein relation and the Stokes paradox for the mobility of a
disk of radius $\sigma$. This system-size dependence of the effective
diffusion coefficient has been verified using steady-state particle
simulations
\citet{DiffusionRenormalization_PRL,DiffusionRenormalization}.  Note
also that \eqref{eq:chi_SE} allows us to validate {\em a posteriori}
the assumption of large separation of time scales between
concentration and momentum diffusion. Specifically, the limiting
equation \eqref{eq:limiting_Strato_Ito} is a good approximation to
\eqref{eq:c_eq_original} if the effective Schmidt number
$\text{Sc}=\nu/\chi_{\text{eff}}=\nu/\left(\chi_{0}+\chi\right)\gg1$.
This is indeed the case in practice for simple liquids and
macromolecular solutions.

%\subsection{\label{sub:Algorithm}Multiscale Numerical Algorithms}

%\section{Is Diffusion Irreversible?}

The measured diffusion coefficients in molecular liquids and
macromolecular solutions closely match the Stokes-Einstein
prediction. This suggests that in realistic fluids diffusive transport
is dominated by advection by the velocity fluctuations,
$\chi\gg\chi_{0}$. Since we know that each realization follows a
strictly reversible dynamics when $\chi_{0}=0$, but that the evolution
of the mean is dissipative even in this case since
$\chi_{\text{eff}}=\chi>0$, it is important to understand the
difference in the behavior of the ensemble mean of the diffusive
mixing process, described by \eqref{eq:dc_dt_mean}, and the behavior
of an individual realization, described by
\eqref{eq:limiting_Strato_Ito}.

%\begin{figure}
%\begin{centering}
%\includegraphics[width=0.475\columnwidth]{SmoothDecay_2D_Lagrangian_coarsenedx2_0000}\hspace{0.2cm}\includegraphics[width=0.475\columnwidth]{SmoothDecay_2D_Lagrangian_coarsenedx2_0004}
%\includegraphics[width=0.475\columnwidth]{SmoothDecay_2D_Lagrangian_0000}\hspace{0.2cm}\includegraphics[width=0.475\columnwidth]{SmoothDecay_2D_Lagrangian_0008}
%\par\end{centering}

\begin{figure}
\begin{centering}
\includegraphics[width=1.0\columnwidth]{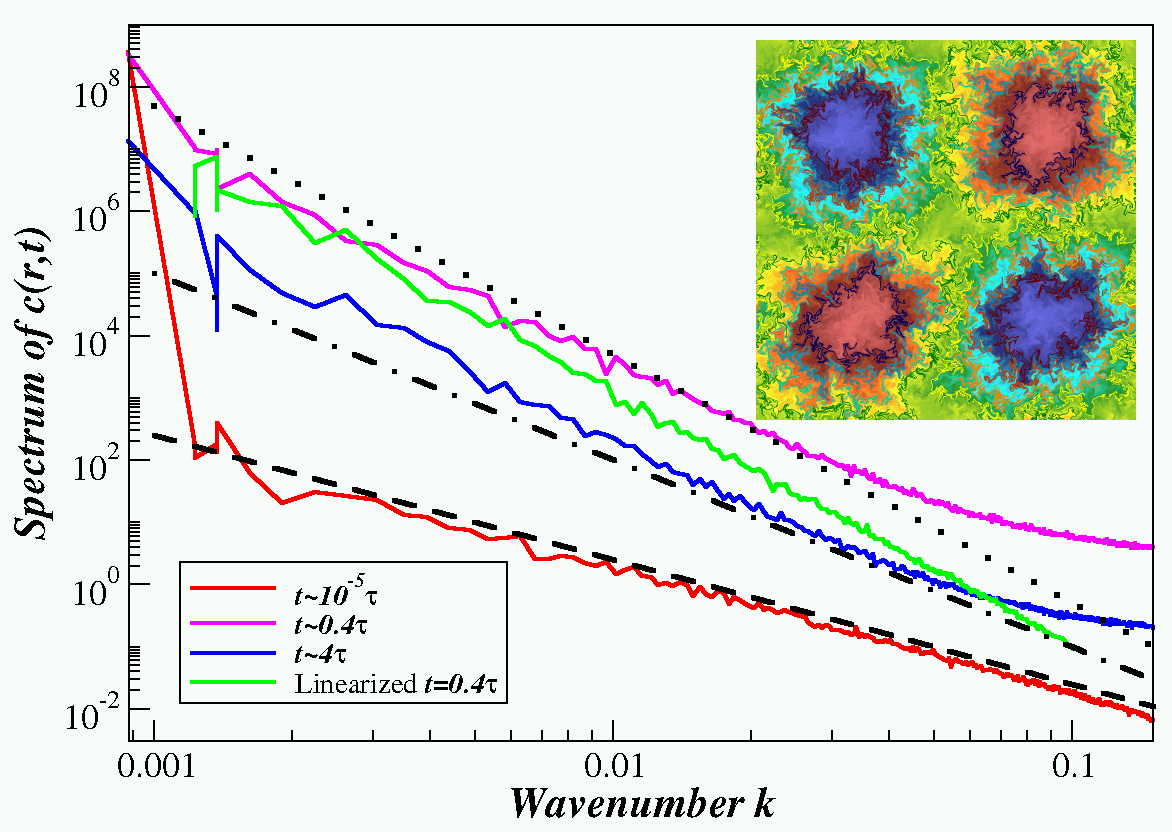} 
\par\end{centering}

\caption{\label{fig:S_k_c} The decay of a single-mode initial
  condition. The inset shows the concentration at a time $t \approx
  2\tau$, as obtained from a Lagrangian simulation with
  $2048^{2}$ tracers and no bare diffusion, along with numerical
  approximations to the contour lines.  The main figure shows the
  power spectrum of an \emph{individual} realization of the
  concentration $c(\V r,t)$ at several times, as obtained using an 
  Eulerian algorithm for solving \eqref{eq:limiting_Strato_Ito}. The power of individual
  modes $\V k$ with nearby $k$ is averaged and the result is shown
  with colored solid lines, while dashed/dotted lines show power laws
  $k^{-2}$, $k^{-3}$ and $k^{-4}$ for comparison. At early times
  $t\ll\tau=\left(2\chi_{\text{eff}}k_{0}^{2}\right)^{-1}$ (red line)
  power is being transferred from mode $\V
  k_{0}\approx2\pi/L\approx10^{-3}$, initially excited to have
  spectral power $p_{\V k_{0}}\approx7\cdot10^{8}$, to the rest of the
  modes, leading to a spectrum $\sim k^{-2}$. At late times
  $t\gtrsim\tau$ (magenta and blue lines), a steadily-decaying shape
  of the spectrum is reached where power transferred from the larger
  scales is dissipated at the small scales via bare diffusion.
  Numerically linearized fluctuating hydrodynamics predicts a spectrum
  $\sim k^{-4}$ (green line) \citet{FluctHydroNonEq_Book}.}
\end{figure}

%\caption{\label{fig:SmoothDecay}
%{\bf Donev: Consider deleting this figure or maybe just making right panel an inset in Fig. 2}
%The decay of a single-mode initial condition
%$c(\V r,0)=\sin\left(2\pi x/L\right)\sin\left(2\pi y/L\right)$ (left)
%to a later time $t \approx 2\tau$ (right), as obtained from a Lagrangian
%simulation with $2048^{2}$ tracers and no bare diffusion. Numerical
%approximations to the contour lines are also shown.}
%\end{figure}
To this end, we resort to numerical experiments using finite-volume \cite{LLNS_Staggered}
Eulerian methods ~\footnote{%
  We checked that the results of these simulations compare well to
  those obtained by integrating the resolved dynamics (\ref{eq:v_eq},
  \ref{eq:c_eq_original}) with
  $S_{c}=\nu/\chi_{\text{eff}}\sim10^{3}-10^{4}$
  \citet{LLNS_Staggered}. The overdamped simulations can reach the same time scales in
  \emph{much} less (by a factor of about $S_{c}$) computational effort
  than the direct numerical simulation because they bypass the need to
  resolve the fast velocity fluctuations.}, 
as well as Lagrangian tracers algorithms~\footnote{%
  The Lagrangian description associated
  with~\eqref{eq:limiting_Strato_Ito} reads
$$
  d\V q=\sum_{k}\V{\phi}_{k}\left(\V q\right)\circ d\mathcal{B}_{k}+\sqrt{2\chi_{0}}\, d\M{\mathcal{B}}_{\V q},
$$
where a \emph{single }realization of the random field
$\sum_{k}\V{\phi}_{k}\circ d\mathcal{B}_{k}$ advects \emph{all} of the
walkers. This induces correlations between their trajectories which
crucially affect the physics of the collective diffusion of the
tracers.}
that we have developed specifically for the purpose of
simulating the limiting dynamics \eqref{eq:limiting_Strato_Ito} at
$\chi_0>0$ and $\chi_0=0$, respectively. Details of these
multiscale numerical methods are given elsewhere \cite{DiffusionJSTAT}. 
Let us consider the temporal decay of a smooth single-mode initial perturbation
$c(\V r,0)=\sin\left(2\pi x/L\right)\sin\left(2\pi y/L\right)$ in two
dimensions.  The ensemble \emph{mean} $\av c$ follows the simple
diffusion equation \eqref{eq:dc_dt_mean}, and therefore remains a
single-mode field with an amplitude decaying as
$\exp\left(-t/\tau\right)$, where
$\tau=\left(2\chi_{\text{eff}}k_{0}^{2}\right)^{-1}$ is a decay time
and $k_{0}$ is the initially excited wavenumber.  In the inset of
Fig.~\ref{fig:S_k_c}, we show a single \emph{instance} (realization)
of the concentration at time $t\approx 2\tau$ when $\chi_0=0$. The figure reveals
characteristic \emph{giant} (long-ranged) fluctuations in particular
realizations of the diffusive process, with the contour lines of the
concentration becoming rougher as time progresses \footnote{%
  We have performed hard-disk molecular dynamics simulation of this
  mixing process and observed the same qualitative behavior seen in
  the inset of Fig. \ref{fig:S_k_c}.}. These enhanced
nonequilibrium fluctuations stem from the development of a power-law
spectrum as the mixing progresses, 
as predicted by linearized fluctuating hydrodynamics \citet{FluctHydroNonEq_Book}.
The evolution of the power spectrum during the diffusive decay is illustrated in
Fig.~\ref{fig:S_k_c}.% , which was obtained using our Eulerian numerical
% method for solving \eqref{eq:limiting_Strato_Ito}.  A small amount of
% bare diffusion was added in the Eulerian simulations to stabilize the
% discretization by dissipating the grid-scale fluctuations.

The conserved quantity $\int\left(c^{2}/2\right)d\V r$ injected via
the initial perturbation away from equilibrium is effectively
dissipated through a mechanism similar to the energy cascade observed
in turbulent flows. Advection transfers power from the large length
scales to the small length scales, \emph{effectively} dissipating the
power injected into the large scales via the initial condition.  A
straightforward calculation that is detailed elsewhere
\citet{DiffusionJSTAT} shows that the total rate at which power is
lost (``dissipated'') from mode $\V k_{0}$ is given by $\V
k_{0}\cdot\M{\chi}\cdot\V k_{0}$. This is exactly the same rate of
dissipation as one would get for ordinary diffusion with diffusion
tensor $\M{\chi}$.  However, in simple diffusion all other modes would
remain unexcited and there would be no giant fluctuations.

At late times of the diffusive decay, $t\sim\tau$, one expects that a
self-similar state will be reached in which the shape of the spectrum
of $c$ does not change as it decays exponentially in time as
$\exp\left(-t/\tau\right)$.  This is indeed what we observe, and the
shape of the decaying spectrum is shown in
Fig. \ref{fig:S_k_c}. Numerically we observe that most of the bare
dissipation occurs at the largest wavenumbers. Note however that the
shape of the spectrum at the large wavenumbers is strongly affected by
discretization artifacts and the presence of (small) bare
diffusion. These numerical grid artifacts can be eliminated by using
the Lagrangian tracer algorithm, which leads to a similar power-law
behavior \cite{DiffusionJSTAT}.

\revision{In the literature, linearized fluctuating hydrodynamics
is frequently used to obtain the steady-state spectrum of fluctuations
\cite{FluctHydroNonEq_Book}. In the limit of large Schmidt numbers,
the standard heuristic approach leads to the additive-noise equation,
\begin{equation}
\partial_{t}\tilde{c}=-\V w\cdot\grad\av c+\grad\cdot\left[\chi_{\text{eff}}\grad\tilde{c}\right],\label{eq:limiting_linearized}
\end{equation}
where $\av c$ is the ensemble mean, which follows (\ref{eq:dc_dt_mean}). Note that
we have not accounted for \emph{equilibrium}
concentration fluctuations in (\ref{eq:limiting_linearized}) since our focus here is on the \emph{nonequilibrium}
fluctuations and we wish to more accurately
measure the power-law spectrum. Equation (\ref{eq:limiting_linearized})
can easily be solved analytically in the Fourier domain when $\grad\av c=\V h$
is a weak externally applied constant gradient to obtain a spectrum
$\left(\V h\cdot\hat{\M{\mathcal{\chi}}}_{\V k}\cdot\V h\right)/\left(\chi_{\text{eff}}k^{2}\right)\sim k^{-4}$
for intermediate wavenumbers. For finite gradients and more realistic boundary
conditions, we can solve (\ref{eq:limiting_linearized}) numerically
with the same algorithm used to solve the full nonlinear equation
(\ref{eq:limiting_Strato_Ito}) by simply reducing the
magnitude of the fluctuations by a large factor and then increasing
the spectrum of the fluctuations by the same factor to obtain
the spectrum of $\tilde{c}$ \cite{LLNS_Staggered}. The result of this numerically-linearized
calculation for the single-mode initial condition is shown in Fig.
\ref{fig:S_k_c} and seen to follow the expected $k^{-4}$ power-law \cite{FluctHydroNonEq_Book}.
This power law is not in a very good agreement with
the spectrum obtained by solving the full nonlinear equation (\ref{eq:limiting_Strato_Ito}), which appears closer to
$k^{-3}$ in the two-dimensional setting we study here.}

If there were only random advection, with no bare diffusion, the
transfer of energy from the coarse to the fine scales would continue
indefinitely, since the dynamics is reversible and there is nothing to
dissipate the power.  However, any features in $c$ at length scales
below molecular scales have no clear physical meaning. In fact,
continuum models are inapplicable at those scales. In typical
experiments, such as FRAP measurements of diffusion coefficients, one
observes the concentration spatially-coarse grained at scales much
larger than the molecular scale.  It is expected that not resolving
(coarse-graining) the microscopic scales will lead to true dissipation
and irreversibility in the coarse-grained dynamics.  Such
coarse-graining can take form of ensemble averaging, or elimination of
slow degrees of freedom. In either case, the loss of knowledge about
the small scales will lead to positive entropy production.

It is reasonable to expect that one can replace the molecular scale
details, or even all details of the dynamics at scales below some
mesoscopic observation scale $\delta$, by effective dissipation. In
particular, we suggest that small-scale details in
(\ref{eq:limiting_Strato_Ito}) can be replaced by a diffusive term
with suitably chosen {\em renormalized} bare coefficient. This
renormalization needs to be carried in such a way that the effective
diffusion coefficient in the equation for the mean remains equal to
$\M{\chi}_{\text{eff}} =\chi_{0}\M{I}+\M{\chi}$.  A partial ensemble
averaging of (\ref{eq:limiting_Strato_Ito}) can be used to achieve
this goal~\citet{DiffusionJSTAT}, and this calculation leads to a
spatially coarse-grained model for diffusion in liquids,
\begin{equation}
  \partial_{t}c_{\delta}=-\V
  w_{\delta}\odot\grad c_{\delta}
  +\grad\cdot\left[\left(\chi_{0}\M{I}+\D{\M{\chi}_{\delta}}\right)
    \grad c_{\delta}\right],
  \label{eq:filtered_c_Strato}
\end{equation}
where $c_\delta = \M{K}_{\delta} \star c$ denotes the concentration filtered at the
mesoscopic scale $\delta$, the white-in-time random velocity $\V
w_{\delta}$ has covariance $\M{K}_{\delta} \star
\M{\mathcal{R}}\star\M{K}_{\delta}^{T}$, and $\chi_0$ is renormalized
by
\begin{equation}
\D{\M{\chi}_{\delta}}(\V r) = \tfrac{1}{2} \left(\M{\mathcal{R}}-
  \M{K}_{\delta} \star
  \M{\mathcal{R}}\star\M{K}_{\delta}^{T}\right)\left(\V r,\V r\right)
\end{equation}
In Ito's form \eqref{eq:filtered_c_Strato} is the same as
\eqref{eq:limiting_Strato_Ito} with $\V w$ replaced by $\V
w_{\delta}$.  Note that the renormalized \emph{bare} diffusion
coefficient
$\M{\chi}_{0}\left(\delta\right)=\chi_{0}\M{I}+\D{\M{\chi}_{\delta}}$
in \eqref{eq:filtered_c_Strato} is nonzero even if $\chi_{0}=0$.  This
true dissipation is a remnant of the unresolved (eliminated) small
scales. However, it is important to stress that
$\M{\chi}_{0}\left(\delta\right)$ is not a material constant, but
rather, depends on the mesoscopic lengthscale $\delta$.

\begin{figure}
\includegraphics[width=1\columnwidth]{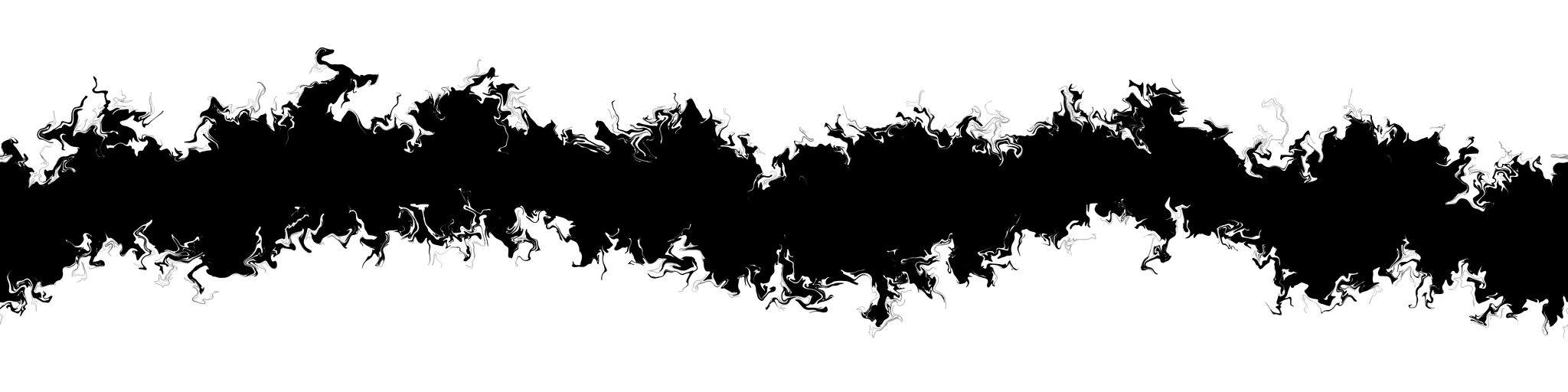}\vspace{0.1cm}

\includegraphics[width=1\columnwidth]{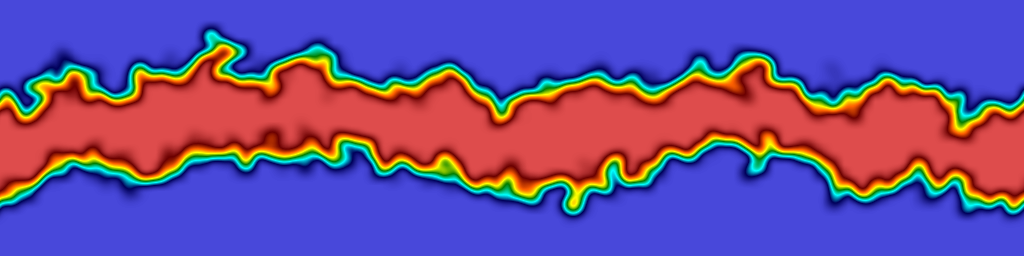}\vspace{0.1cm}
 
\includegraphics[width=1\columnwidth]{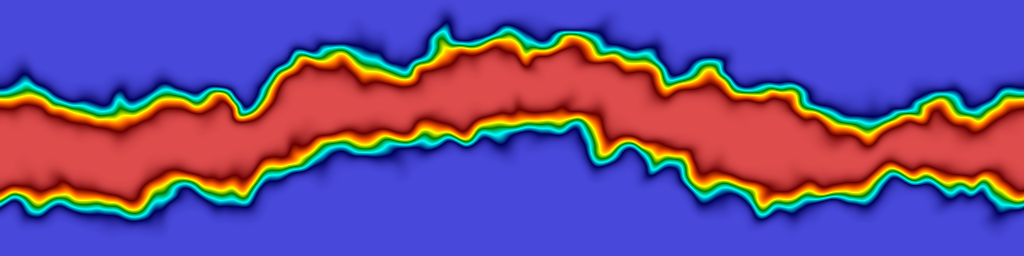} 

\caption{\label{fig:LagrangianEulerian} \emph{Top panel}: A snapshot
  of the concentration $c$ for diffusive mixing of two miscible fluids
  in the absence of bare diffusion, starting from concentration being
  unity (black) in a horizontal stripe occupying one third of the
  periodic domain, and zero (white) elsewhere. The top and bottom
  interface are represented with about half a million Lagrangian
  tracers each. \revision{\emph{Middle panel}:
The spatially-coarse grained concentration $c{}_{\delta}$
obtained by blurring the top panel using a Gaussian filter with standard
deviation $\delta=3\sigma$.
\emph{Bottom panel}: An independent snapshot of the
  spatially coarse-grained concentration $c_{\delta}$ at the same point in time as
  the top panel, obtained by solving
  \eqref{eq:filtered_c_Strato} with an Eulerian method using a grid of
  $2048\times512$ finite-volume cells. A Gaussian filter of width
  $\delta=3\sigma$ is used to filter the discrete velocity. The
effective diffusion coefficient $\chi_{\text{eff}}$ is the same as in the top panel.}}
\end{figure}

To test~\eqref{eq:filtered_c_Strato}, consider diffusive mixing
between two initially phase-separated fluids in two dimensions with
periodic boundary conditions.  We start with concentration $c=1$ in a
thin horizontal stripe, $c=0$ everywhere.  This could, for example,
model a stripe in a FRAP experiment in which a laser beam combined
with a diffraction grating is used to create a striped pattern of
fluorescently labeled tracers at $t=0$.  In
Fig. \ref{fig:LagrangianEulerian} we show snapshots of the
concentration field at a later time, for $\delta=0$ (no
coarse-graining) in the top panel, and $\delta=3\sigma$ in the middle and bottom
panels. Specifically, in the middle panel we show 
the spatially smoothed concentration ${\M{K}_{\delta}}\star c$.
For comparison, in the bottom panel we show an instance of the solution
of the proposed coarse-grained diffusion equation \eqref{eq:filtered_c_Strato}.  
Since $\V u$ and $\V w$ are spatially-smooth velocity fields,
advection by these fields leads to behavior qualitatively different
from diffusion when $\chi_0=0$. Specifically, if the initial
concentration $c\left(\V r,0\right)$ has a sharp interface, this
interface will remain sharp at all times, even if it becomes very
rough, at \emph{all} times, in \emph{every} realization. Therefore,
the top panel in the figure is black and white.  In the presence of
true (bare) dissipation, the interface between the two fluids does not
remain sharp, and a range of concentrations $0 \leq c \leq 1$ appears
for $t>0$.  Therefore, the middle and bottom panels in the figure show
a spectrum of colors.

\revision{
In large three dimensional systems, when the spatial coarse-graining
is performed at macroscopic scales $\delta\gg\sigma$, it is expected that 
\eqref{eq:filtered_c_Strato} will converge in some sense to the linearized fluctuating
hydrodynamics equation (\ref{eq:limiting_linearized}), as suggested by renormalization arguments \cite{DiffusionRenormalization_I}.
While we are not aware of mathematical tools to prove this type of statement, 
a plausible argument goes as follows.
In three dimensions, as $\delta\rightarrow\infty$, the renormalization
of the diffusion coefficient approaches the Stokes-Einstein value,
$\D{\M{\chi}_{\delta}}\rightarrow\M{\chi}$, and the stochastic term
$\V w_{\delta}\odot\grad c{}_{\delta}$ becomes negligible because
most of the power (spectral intensity) in the the random advection
velocity $\V w_{\delta}$ is removed by the filtering (since the spectrum
of $\V w$ decays like $k^{-2}$, in three dimensions the power distribution
is independent of $k$). Therefore the noise will become ``weak''
in a suitable sense and the fluctuations can be linearized around
the mean. This is not true in two dimensions, where large scale features
in $\V w_{\delta}$ give the dominant contribution to the effective
diffusion and contain the majority of the spectral power of $\V w$
(in two dimensions the power distribution decays like $k^{-1}$). Therefore,
linearization is certainly not appropriate in two dimensions even if $\delta\gg\sigma$.
Thin films may exhibit an intermediate behavior
depending on the scale of observation relative to the thickness of
the thin film \cite{ThinFilm_Smectic}.}

Summing up, in both two and three dimensions the behavior of mixing
processes in liquids cannot be described by Fick's law at mesoscopic
scales. One must include random advection by the mesoscopic scales of
the velocity fluctuations in order to reproduce not just the behavior
of the mean but also the long-range correlated fluctuations observed
in individual realizations.  
This emphasizes the crucial distinction between the self-diffusion of
{\em individual} tracers and the {\em collective} diffusion of many hydrodynamically-correlated tracers. 
% These giant fluctuations and the
% diffusion renormalization depend sensitively on the spectrum of the
% velocity fluctuations, which is affected by boundary conditions
% (confinement)
% \citet{FractalDiffusion_Microgravity,Nanopore_Fluctuations,ThinFilm_Smectic,DiffusionRenormalization}.
The traditional Fick's diffusion constant is only meaningful under
special conditions (e.g., large three-dimensional bulk systems
observed at macroscopic scales) which may not in fact be satisfied in
experiments aimed to measure ``the'' diffusion coefficient. A length
scale of observation (coarse-graining) must be attached to the
diffusion coefficient value in order to make it a ``material
constant'' that can be used in a predictive model of diffusive
transport \citet{DiffusionRenormalization}.
Furthermore, the measured diffusion coefficient is strongly affected by
boundary conditions (confinement) \citet{Nanopore_Fluctuations,ThinFilm_Smectic,DiffusionRenormalization_PRL}.

We hope that these results
will spur interest in designing experiments that carefully examine
diffusion at a broad range of length scales.
Existing experiments
have been able to measure concentration fluctuations across a wide
range of lenghtscales transverse to the gradient, but fluctuations are
averaged longitudinally over essentially macroscopic scales (thickness
of the sample) \citet{GiantFluctuations_Nature,GiantFluctuations_Cannell,FractalDiffusion_Microgravity}.
While FRAP experiments routinely look at diffusion at micrometer
scales, we are not aware of any work that has even attempted to
account for the effect of thermal fluctuations. Giant fluctuations are
expected to be more easily observed in thin liquid films due to the
quasi-two dimensional geometry \citet{ThinFilm_Smectic,LiquidCrystalFilms}.
\revision{In the future we will
consider extensions of our approach to multispecies liquid mixtures. Such extensions
are expected to lead to a better understanding of the physics of diffusion in fluid mixtures,
including a generalized Stokes-Einstein relation for inter-diffusion coefficients in
dilute multispecies solutions.}

\newpage

\begin{acknowledgments}
We would like to acknowledge Florencio Balboa Usabiaga and Andreas Klockner for their help in developing a GPU implementation of the numerical methods, and Leslie Greengard for advice on the non-uniform FFT algorithm. We are grateful to Alberto Vailati, Alejandro Garcia, John Bell, Sascha Hilgenfeldt, Mike Cates and Ranojoy Adhikari for their insightful comments. 
A. D. was supported in part by the NSF grant DMS-1115341 and the DOE
Early Career award DE-SC0008271. T. F.was supported in part by the DOE
CSGF grant DE-FG02-97ER25308.  E. V.-E. was supported by the DOE ASCR
grant DE-FG02-88ER25053, the NSF grant DMS07-08140, and the ONR grant
N00014-11-1-0345.
\end{acknowledgments}

%\newpage

%\bibliographystyle{apsrev}
%\bibliographystyle{eplbib}
%\bibliography{References,10_home_donev_Papers_DiffusiveMixing_GarciaGeneralBibFile,11_home_donev_Papers_DiffusiveMixing_MScaleProp}

\end{document}